# Monitoring transition: expected night sky brightness trends in different photometric bands

**Salvador Bará,**[1,*] **Iago Rigueiro,**[1] **and Raul C. Lima**[2,3]

[1] *Departamento de Física Aplicada, Universidade de Santiago de Compostela, 15782 Santiago de Compostela, Galicia.*

[2] *Physics, Escola Superior de Saúde, Politécnico do Porto, Portugal.*

[3] *CITEUC – Centre for Earth and Space Research, University of Coimbra, Portugal.*

[*] salva.bara@usc.gal

**Abstract**

Several light pollution indicators are commonly used to monitor the effects of the transition from outdoor lighting systems based on traditional gas-discharge lamps to solid-state light sources. In this work we analyze a subset of these indicators, including the artificial zenithal night sky brightness in the visual photopic and scotopic bands, the brightness in the specific photometric band of the widely used Sky Quality Meter (SQM), and the top-of-atmosphere radiance detected by the VIIRS-DNB radiometer onboard the satellite Suomi-NPP. Using a single-scattering approximation in a layered atmosphere we quantitatively show that, depending on the transition scenarios, these indicators may show different, even opposite behaviors. This is mainly due to the combined effects of the changes in the sources' spectra and angular radiation patterns, the wavelength-dependent atmospheric propagation processes and the differences in the detector spectral sensitivity bands. It is suggested that the possible presence of this differential behavior should be taken into account when evaluating light pollution indicator datasets for assessing the outcomes of public policy decisions regarding the upgrading of outdoor lighting systems.

## 1. Introduction

The accelerated pace of substitution of traditional outdoor gas-discharge lamps by solid-state lighting technology sources (SSL) raises some relevant questions regarding its expected consequences on the light pollution levels recorded in the neighborhood of the intervention areas, as well as in places located far away from them. Several key radiometric variables are substantially modified when traditional lamps are replaced by LEDs, including the spectral composition of the light, the spatial and angular distribution of the spectral radiance emitted by the luminaires, and, possibly, the average illumination levels. It could well be expected that these changes will not be neutral regarding the evolution of different indicators associated with light pollution, among them the anthropogenic night sky brightness.

Due to the negative consequences of light pollution for the preservation of ecosystems [1-9], ground-based astrophysics [10-13], humankind's intangible heritage [14-15], and potentially, public health [16-20], there is nowadays an intense research effort to accurately monitor the light pollution trends worldwide. Two main and complementary approaches are commonly used: (a) tracking the changes of the emitted artificial radiance by means of ground-based measurements [21-23], airborne surveys [24-25], or remote sensing from Earth orbiting platforms [26-31], and (b) monitoring the evolution of the night sky brightness using specific detectors [32-39], CCD and general purpose DSLR cameras [40-43], and even by means of comprehensive citizen science observations [44]. Different trends have been reported, sometimes apparently at odds with each other, including a sustained increase of the radiance and overall lit surface around the world, at an estimated 2% yearly rate [45], and a tendency to darker readings in the long-term series recorded by ground-based night sky brightness detectors [46-47], whose progressive aging has not been discarded as a potential cause.

It shall be kept in mind that these measurements are made in different spectral bands, and after the light has undergone different wavelength-selective propagation processes. It can be expected that the detected evolution trends may depend on the specific observation channels, so the possibility of observing seemingly diverging results regarding the sign of these trends shall not be discarded. A convincing case has been made in a recent paper by Sánchez de Miguel et al. [48], who have shown that, under appropriate circumstances, the scotopic and photopic luminances associated with the close-range direct-viewing of lamps in the transition from high-pressure sodium lamps (HPS) to high correlated color temperature (CCT) LEDs may show an opposite behavior.

In this paper we further generalize Sánchez de Miguel et al. results [48] to the more frequent case of measurements made after the light emitted by the luminaires has interacted with the surrounding surfaces (pavements and façades) and has propagated throughout the atmosphere, undergoing absorption and scattering in the molecular and aerosol atmospheric constituents, before being detected by the observer. Our goal is to model the expected evolution of the night sky brightness, measured in different photometric channels and at different distances from the sources, along the transition process of substitution of traditional gas-discharge lamps by LEDs. To that end we consider several photometric detection bands of practical interest, including the photopic [49] and scotopic [50] spectral eye sensitivities, the specific photometric band of the Sky Quality Meter (Unihedron, Calgary, Canada) [51-52] used in several night sky brightness measurement networks across the world, and from which extensive dataset records exist [32-39], and the Day-Night Band of the Visible and Infrared Imaging Radiometer Suite (VIIRS-DNB) onboard the Suomi-NPP satellite [53-54]. Our approach is intentionally kept simple, considering only direct radiance propagation, averaged ground reflections and single-scattering processes, but it is expected to be reasonably comprehensive as to be useful for describing the principal traits of the expected light pollution evolution trends.

Three main scenarios for the transition process from High Pressure Sodium (HPS) to Light Emitting Diode (LED) outdoor lighting are analyzed in this paper: (a) 'Business as usual', in which the direct emissions to the upper hemisphere, the utlance of the installations and the average lighting levels are kept the same before and after the remodeling, (b) the 'zero direct upward flux', in which the average ground illumination levels are maintained but the direct upward emissions of the luminaires are set to zero, with the corresponding increase of the utlance, and (c) the 'lighting reduction' approach, in which the zero direct emissions to the upper hemisphere of option (b) are accompanied by an additional reduction of the average ground lighting levels, either by adopting permanent lower illuminances or by means of appropriate dimming curfews.

## 2. Methods

*2.1 Basic light propagation processes and photometric bands*

A non-negligible fraction of the light emitted by outdoor lighting systems propagates toward the sky, directly from the luminaires or after reflection off the surrounding surfaces. Part of this light may reach observers located far away from the sources, possibly after experiencing one or more individual scattering events. For a ground-based observer this gives rise to an increase of the night sky brightness in any direction of the sky, in comparison with what would be observed if artificial lights were completely absent. Radiometers onboard Earth orbiting platforms, in turn, will reveal the presence of artificial light sources both from the direct source radiance received through the line-of-sight, and from the radiance detected in nadir angles close to the source position, due to atmospheric scattering. Since all these phenomena are wavelength dependent, the signal recorded by each detector type is expected to vary in a specific way along the lamp transition period, according to the progressive changes in the sources' spectra, the spectral weighting produced by the physical processes acting on the way from the source to the detector, and the spectral sensitivity of the detector itself.

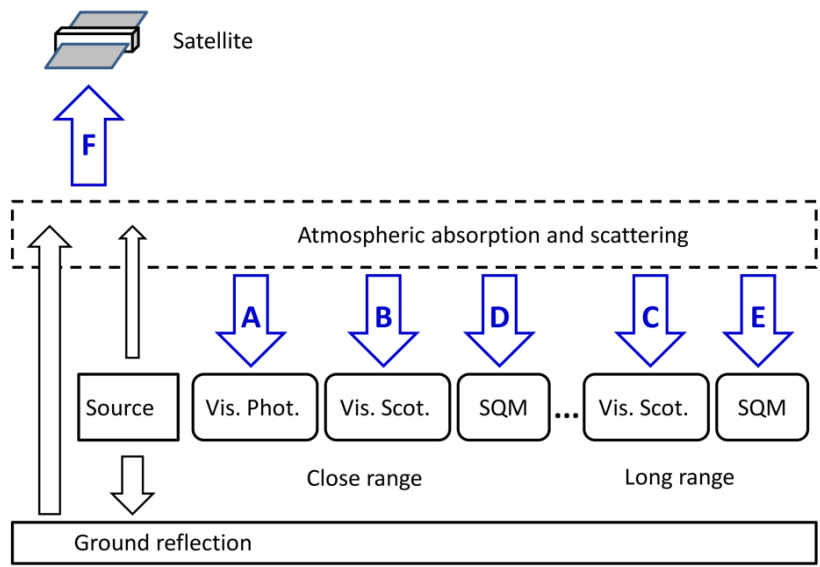

**Figure 1.** Basic light propagation processes and detection channels.

Figure 1 depicts the basic light propagation processes and detection bands included in this work. Regarding sources, we consider both their direct emissions above the horizontal and the upward emissions resulting from Lambertian reflections off the pavements below them. Atmospheric propagation under the single-scattering approximation will be used to compute the artificial spectral radiance in arbitrary directions of the sky at different distances from the light sources. From this radiance the signal detected by the human visual system under photopic adaptation in the neighborhood (< 0.5 km) of the sources (Figure 1, label A) can be easily computed, as well as the visual scotopic signal (B,C) and the SQM one (D,E), respectively, for the whole distance range here considered (from 0.1 km to 100 km away from the sources). The VIIRS-DNB signal (F) can be determined from the top-of-atmosphere (TOA) spectral radiance and the spectral sensitivity of the VIIRS-DNB radiometer. Figure 2 shows the spectral sensitivity bands of these detectors [49-50,52,54].

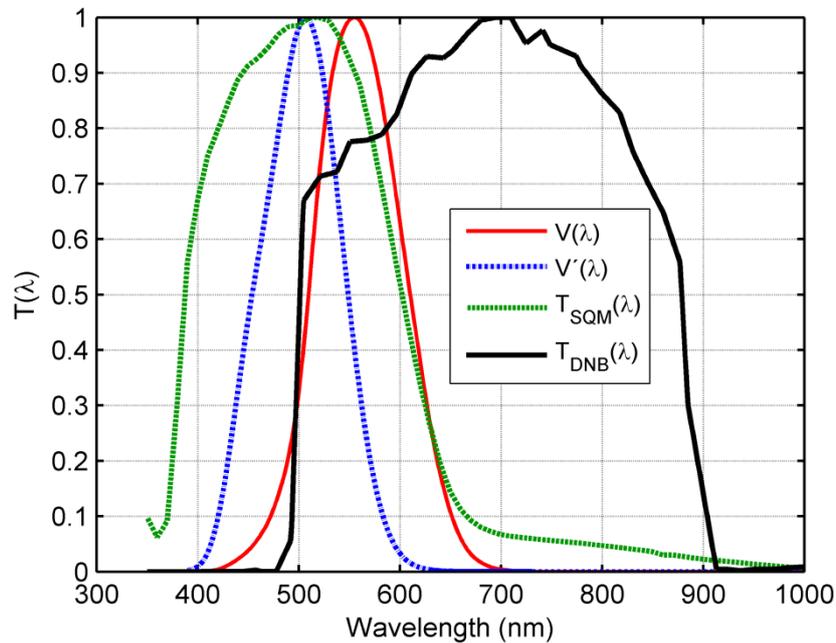

**Figure 2.** Spectral sensitivity bands of the detectors included in this work: CIE photopic, $V(\lambda)$; CIE scotopic, $V'(\lambda)$; Sky Quality Meter, $T_{SQM}(\lambda)$; and VIIRS-DNB, $T_{DNB}(\lambda)$.

The spectral weighting functions associated with the propagation and scattering in the atmosphere for each observation channel (Figure 1) depend on the angular radiation pattern of the sources, the particular atmospheric conditions, and the direction of observation. For the purposes of this work these spectral weighting functions will be calculated using a single-scattering approximation based on the model developed by Kocifaj (2007) [55], which is briefly described in the Appendix and whose results are applied in subsection 2.3 below.

*2.2 The transition process: modeling the time course of the artificial light emissions*

In order to model the time evolution of the artificial light emissions in each transition scenario it is advisable to reformulate this problem in terms of several commonly used lighting parameters, as e.g. the average illuminance levels, the total surface intended to be lit, the utilance of the installations, and the upward flux fraction of the luminaires. The introduction and use of these lighting engineering terms does not change the physics of the problem, yet it makes somewhat easier to describe the time evolution of the emissions (overall spectral flux and spectral radiance) for each transition approach.

To that end, let $\Phi'_i(\lambda)$ be the spectral radiant flux (W·nm$^{-1}$) required to produce a unit illumination level (1 lx) on a unit target area (1 m$^2$), that is, a useful luminous flux of 1 lumen, by means of a given luminaire of type *i*. By "target area" we denote the area that is explicitly intended to be lit, in order to facilitate human activity, as opposed to the total area actually lit by this type of luminaire, that may be larger due to the existence of spilled light. The useful spectral flux required to lit a target area of $M$ square meters with a baseline level of $E$ lx using this kind of luminaire is then $ME\Phi'_i(\lambda)$ W·nm$^{-1}$. The total spectral flux emitted in these conditions is $ME\Phi'_i(\lambda)\eta_i^{-1}$ W·nm$^{-1}$, where $\eta_i$ is the average utilance of the installations that use *i*-type luminaires, that is, the ratio of their useful flux to their total emitted flux.

A fraction $U_i$ (the upward flux fraction) of this spectral flux will be emitted directly toward the sky, totaling $MEU_i\Phi'_i(\lambda)\eta_i^{-1}$ W·nm$^{-1}$. By "directly" we mean "not including ground reflections". The remaining flux, $ME(1-U_i)\Phi'_i(\lambda)\eta_i^{-1}$, will then be emitted toward the ground, lighting both the target area and perhaps neighboring zones where light spill gives rise to light intrusion. Overall, a fraction $1-U_i-\eta_i$ of the total emitted flux arrives to ground areas that in principle were not intended to be lit.

Any target area at the district or city level can be conveniently lit using a combination of different types of luminaires, $i=1,...,N$, whose relative number may vary over time, due e.g. to the replacement of HPS lamps by LEDs. To account for this progressive change, let us denote by $\gamma_i(t)$ the fraction of the target area lit at time $t$ by luminaires of type $i$, characterized by their basic spectral power distribution $\Phi'_i(\lambda)$, utilance $\eta_i$, and upward flux fraction $U_i$. Let us allow for changes of the total target surface (e.g. area increases for new urban developments) by using a multiplicative factor $\mu(t)$ acting on $M$, and for changes in the average illuminance levels (e.g. to accommodate the installations to new lighting regulations, including dimming curfews) through a factor $\varepsilon_i(t)$ acting on $E$. Then, the overall spectral flux directed below the horizontal and illuminating the ground is:

$$\Phi_g(\lambda, t) = \sum_{i=1}^{N} \gamma_i(t)\mu(t)M\varepsilon_i(t)E\frac{\Phi'_i(\lambda)}{\eta_i}(1 - U_i), \qquad (1)$$

giving rise to an average ground spectral irradiance equal to:

$$E_g(\lambda, t) = \frac{\Phi_g(\lambda, t)}{A} = \frac{1}{A}\sum_{i=1}^{N} \gamma_i(t)\mu(t)M\varepsilon_i(t)E\frac{\Phi'_i(\lambda)}{\eta_i}(1 - U_i), \qquad (2)$$

where $A$ is the total ground area (m²) lit by the luminaires, which is generally larger than $\mu(t)M$ because the utilances are generally smaller than $1 - U_i$. The definite value of $A$ depends on the characteristic emitting pattern of each luminaire type and the way they are installed. The reflected radiance $L_r(\lambda, t)$ can be calculated as the product of the incident $E_g(\lambda, t)$ by the ground bidirectional reflectance distribution function (*BRDF*). For Lambertian surfaces the *BRDF* does not depend on the angles of incidence and reflection, and is given by $R_{BRDF}(\lambda) = \pi^{-1}\rho(\lambda)$, where $\rho(\lambda)$ is the spectral ground reflectance. In that case the reflected radiance is constant in all directions, equal to

$$L_r(\lambda, t) = \frac{\rho(\lambda)}{\pi A}\sum_{i=1}^{N} \gamma_i(t)\mu(t)M\varepsilon_i(t)E\frac{\Phi'_i(\lambda)}{\eta_i}(1 - U_i), \qquad (3)$$

and the total reflected flux is

$$\Phi_r(\lambda, t) = \rho(\lambda)\sum_{i=1}^{N} \gamma_i(t)\mu(t)M\varepsilon_i(t)E\frac{\Phi'_i(\lambda)}{\eta_i}(1 - U_i). \qquad (4)$$

On the other hand, the total spectral flux emitted by the luminaires directly above the horizontal is

$$\Phi_u(\lambda, t) = \sum_{i=1}^{N} \gamma_i(t)\mu(t)M\varepsilon_i(t)E\frac{\Phi'_i(\lambda)}{\eta_i}U_i. \qquad (5)$$

The radiance $L_u(\lambda, t; \boldsymbol{\omega}')$ associated with this upward flux, averaged over the total area lit, can be determined by dividing Equation (5) by $A$ and multiplying each term by its angular radiance distribution $G_i(\boldsymbol{\omega}')$ such that

$$L_u(\lambda, t; \boldsymbol{\omega}') = \frac{1}{A}\sum_{i=1}^{N} \gamma_i(t)\mu(t)M\,\varepsilon_i(t)E\frac{\Phi'_i(\lambda)}{\eta_i}U_i\,G_i(\boldsymbol{\omega}'), \qquad (6)$$

where $\boldsymbol{\omega}' = (z_0, \phi_0)$, being $z_0$ the zenith angle and $\phi_0$ the azimuth in the source's reference frame. Note that this $L_u(\lambda, t; \boldsymbol{\omega}')$ is the source radiance averaged over the territory of area $A$, not the average radiance of the lamps themselves. The latter is considerably higher since the sum of the areas of the radiating surfaces of the lamps is much smaller than $A$. In this work we will consider sources with azimuthally symmetric $G_i(\boldsymbol{\omega}')$ functions limited to relatively small angles above the horizontal (large zenithal angles), e.g. $G_i(\boldsymbol{\omega}') = 0$ for $z_0 < 70°$.

*2.3 Light propagation through the atmosphere*

The description of the light propagation through the atmosphere can be done using radiative transfer models with different levels of complexity [56]. For the purposes of this work we will adopt the single-scattering model developed by Kocifaj [55], applying it to source radiances given by the Lambertian ground reflection (Equation 3) and the area-averaged radiance emitted directly above the horizontal (Equation 6).

As shown in the Appendix, the elementary scattered spectral radiance $\mathrm{d}L_R(\lambda, t; \boldsymbol{\omega})$ reaching the observer from the direction $\boldsymbol{\omega} = (z, \varphi)$, where $z$ is the zenith angle and $\varphi$ is the azimuth in the observer's reference frame, due to the spectral flux density $\mathrm{d}\Phi_T(\lambda, t)$ emitted at time $t$ from a small area element $\mathrm{d}A_0$ of the source region located at a distance $D$ from the observer can be formally expressed as

$$\mathrm{d}L_R(\lambda, t; \boldsymbol{\omega}) = K(\lambda; \boldsymbol{\omega}; D)\mathrm{d}\Phi_T(\lambda, t), \qquad (7)$$

where the function $K(\lambda; \boldsymbol{\omega}; D)$ depends on the angular radiance distribution of the sources and on the atmospheric conditions, in addition to the explicit variables indicated in its argument. The total spectral radiance reaching the observer, $dL_R(\lambda, t; \boldsymbol{\omega})$, can be obtained by integrating Equation (7) over the area of the source region. Since we are interested in describing general trends we will assume for the purposes of this work that, for each type of source and atmospheric conditions, the function $K(\lambda; \boldsymbol{\omega}; D)$ in Equation (7) can be considered approximately constant for small emitting areas of size $<< D$. Additionally, since we are considering two different types of source contributions with specific angular radiating patterns (the ground Lambertian reflection, henceforth denoted by the subindex '$r$', plus the direct radiance emitted by the luminaires in directions above the horizontal, henceforth denoted by the subindex '$u$'), two radiant flux contributions with their specific $K$ functions have to be taken into account. Under these assumptions the radiance reaching the observer can be described by:

$$L_R(\lambda, t; D; \boldsymbol{\omega}) = K_r(\lambda; \boldsymbol{\omega}; D)\Phi_r(\lambda, t) + K_u(\lambda; \boldsymbol{\omega}; D)\Phi_u(\lambda, t) \qquad (8)$$

The particular case $\boldsymbol{\omega} = (0,0)$, corresponding to the zenithal night sky brightness observations, will be analyzed in this work.

On the other hand, the radiance reaching the top of the atmosphere along a vertical path over the sources, $L_{TOA}(\lambda, t)$, can be calculated from the ground emitted radiance taking into account the extinction due to absorption and scattering by the air molecules and aerosols. In terms of $\tau_0^{(M)}(\lambda)$ and $\tau_0^{(A)}(\lambda)$, the molecular and aerosol optical depths, respectively, the TOA radiance can be expressed as:

$$L_{TOA}(\lambda, t) = \exp\left\{-\left[\tau_0^{(M)}(\lambda) + \tau_0^{(A)}(\lambda)\right]\right\} L_r(\lambda, t), \qquad (9)$$

where, for the simplified conditions of our calculation, we consider that the observation is made when the satellite is on the vertical of the sources (hence through a unit airmass), and that only the Lambertian component $L_r(\lambda, t)$ of Equation (3) is a relevant contributor, since the direct upward flux of the luminaires $L_u(\lambda, t; \boldsymbol{\omega}')$ is limited to large zenithal angles and hence it is not directly detected by the on-orbit radiometer when observing in the nadir direction.

## 2.4 Light pollution indicators

The signal $S(t)$ provided by a detector with spectral sensitivity $T(\lambda)$ when measuring a field-of-view averaged radiance $L(\lambda, t)$ is given by

$$S(t) = k \int_\lambda T(\lambda) L(\lambda, t)\, d\lambda,  \qquad (10)$$

where $k$ is a scale constant. The functions corresponding to the channels considered in this work are indicated in Table 1.

**Table 1.** Spectral functions for the different detection channels

| Channel | $T(\lambda)$ | $L(\lambda, t)$ | $k$ |
|---|---|---|---|
| Zenithal sky brightness, CIE photopic | $V(\lambda)$ | $L_R(\lambda, t; D; \boldsymbol{\omega} = \mathbf{0})$ | 683 lm/W |
| Zenithal sky brightness, CIE scotopic | $V'(\lambda)$ | $L_R(\lambda, t; D; \boldsymbol{\omega} = \mathbf{0})$ | 1700 lm/W |
| Zenithal sky brightness, SQM | $T_{SQM}(\lambda)$ | $L_R(\lambda, t; D; \boldsymbol{\omega} = \mathbf{0})$ | 1.0 |
| Nadir radiance, VIIRS-DNB | $T_{DNB}(\lambda)$ | $L_{TOA}(\lambda, t)$ | 1.0 |

A convenient representation for the time course of these signals can be obtained by using a Pogson-like [57] traditional astronomical magnitude notation, whereby the change of magnitudes of the signal in any generic channel $\alpha$ at time $t$ with respect to its initial value at time $t_0$ is given by

$$\Delta m_\alpha(t) = -2.5 \log_{10}\left[\frac{S_\alpha(t)}{S_\alpha(t_0)}\right]. \qquad (11)$$

## 3. Results

### 3.1. Lamp spectra, ground reflectance, atmospheric conditions, and transition scenarios

We applied the model described in section 2 to a transition process consisting on the replacement of High Pressure Sodium lamps (HPS, CCT≈2200 K) by white LEDs of CCT 4000 K. The spectra of two typical examples of both type of sources are displayed in Figure 3. The reason for choosing this type of LED CCT, which should not be taken as an endorsement by the authors of this article, is due to the fact that they are currently the most commonly used type in Galicia to replace older technologies.

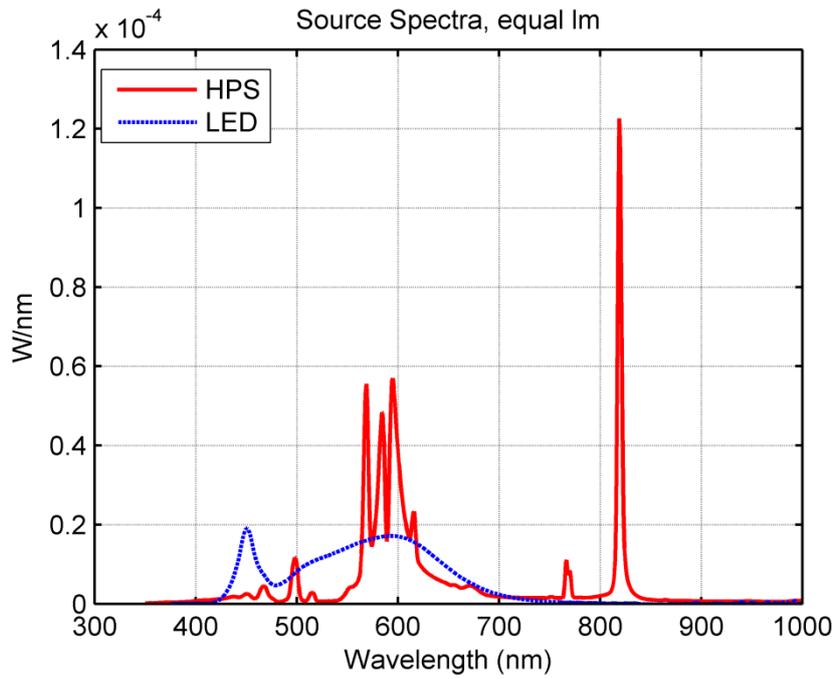

**Figure 3.** Spectra of HPS and 4000 K LED lamps

An average pavement reflectance function has been used in the calculations, combining with weights 1:1:1 two concrete plus one asphalt samples from the USGS Spectral Library [58] as shown in Figure 4.

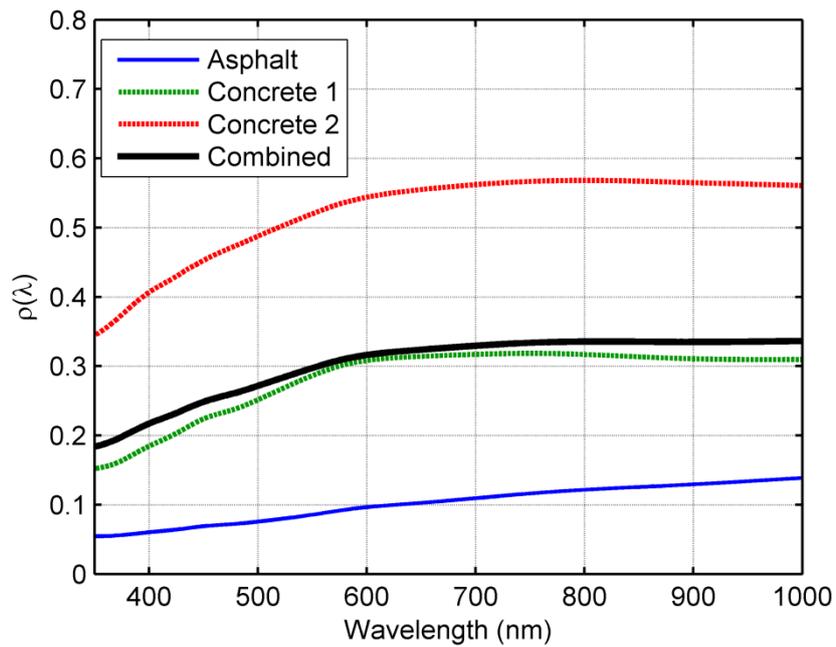

**Figure 4.** Spectral reflectances of concrete and asphalt pavements, and combined average value used in this work.

The atmospheric parameters required for computing the light propagation throughout the atmosphere are summarized in Table 2. The aerosol ones correspond to the 2012-2015 average conditions in A Coruña, as determined by the AERONET station located at this city of NW Galicia [59].

**Table 2.** Atmospheric parameters used in the model

| Parameter | Symbol | Value |
| --- | --- | --- |
| Molecular scale height | $h_m$ | 8.0 km |
| Aerosol scale height | $h_a$ | (1/0.65) km |
| Aerosol albedo | $\Omega^{(A)}$ | 0.92 |
| Aerosol asymmetry parameter | $g$ | 0.65 |
| Reference wavelength | $\lambda_0$ | 442 nm |
| Aerosol optical depth at $\lambda_0$ | $\tau_0^{(A)}(\lambda_0)$ | 0.25 |
| Angstrom exponent aerosols | $\alpha$ | 1.2 |

The atmospheric propagation spectral functions $K_r(\lambda; \boldsymbol{\omega}; D)$ and $K_u(\lambda; \boldsymbol{\omega}; D)$ for zenithal observation ($\boldsymbol{\omega} = \boldsymbol{0}$), computed as indicated in the Appendix, are displayed in Figure 5. For the calculation of $K_u(\lambda; \boldsymbol{\omega}; D)$ we assumed sources with uniform radiance within their angular emission range, i.e. $G_i(\boldsymbol{\omega}') = 1$ for $70° < z_0 \leq 90°$, and $G_i(\boldsymbol{\omega}') = 0$ otherwise.

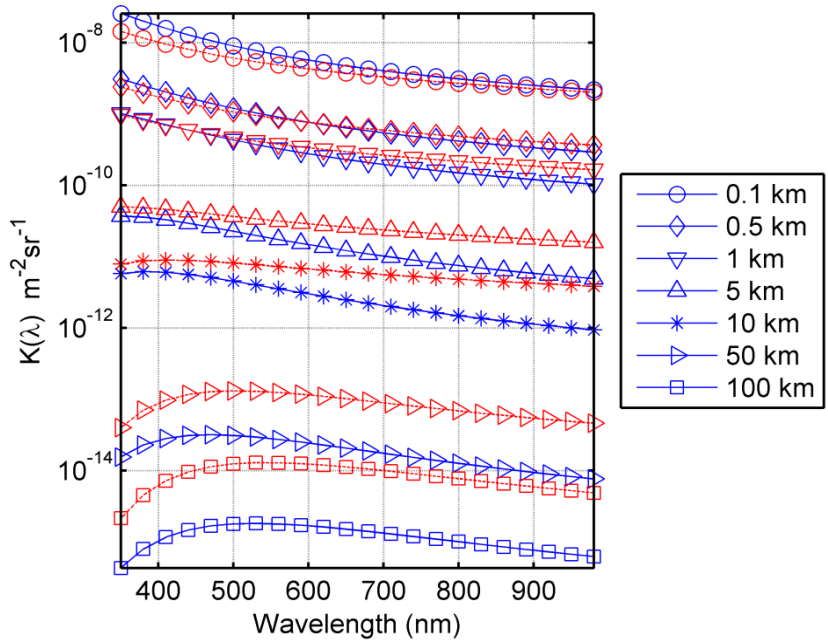

**Figure 5.** Atmospheric propagation spectral functions for zenithal observation $K_r(\lambda; \boldsymbol{0}; D)$ (blue, full line) and $K_u(\lambda; \boldsymbol{0}; D)$ (red, dotted line), according to Equation (8).

Three basic transition scenarios were considered, according to the choices of some key photometric parameters as, e.g. the utilance of the installations or the maximum allowed upward flux fraction. In the first one ('Business as usual') LED lights are installed maintaining the same photometric parameters of the former HPS luminaires. In the second one ('zero direct upward flux'), the direct emissions to the upper hemisphere are set to zero, with the corresponding increase in the utilance of the installations. In the third one ('Lighting reduction'), the average flux emitted along the night is reduced, by means of a reduction of the baseline lighting levels possibly combined with appropriate dimming curfews. The photometric parameters corresponding to these three scenarios are summarized in Table 3.

**Table 3.** Transition scenarios

| Parameter / scenario | HPS | LED | | |
|---|---|---|---|---|
| | Starting point | Business as usual | Zero direct upward flux | Lighting reduction |
| Upward flux fraction ($U_i$) | 0.15 | 0.15 | 0.0 | 0.0 |
| Utilance ($\eta_i$) | 0.4 | 0.4 | 0.6 | 0.6 |
| Light level reduction ($\varepsilon_i$) | 1.0 | 1.0 | 1.0 | 0.85 |
| Surface factor ($\mu$) | 1.0 | 1.0 | 1.0 | 1.0 |

## 3.2 Expected evolution of the light pollution indicators

The expected evolution of the artificial zenithal night sky brightness expressed in the negative logarithmic scale of Equation (11) for the visual photopic, visual scotopic, and SQM photometric bands under the three assumed scenarios is displayed in Figure 6. A selection of these curves is also plotted in Figure 7, combined with the expected signal detected in the VIIRS-DNB band.

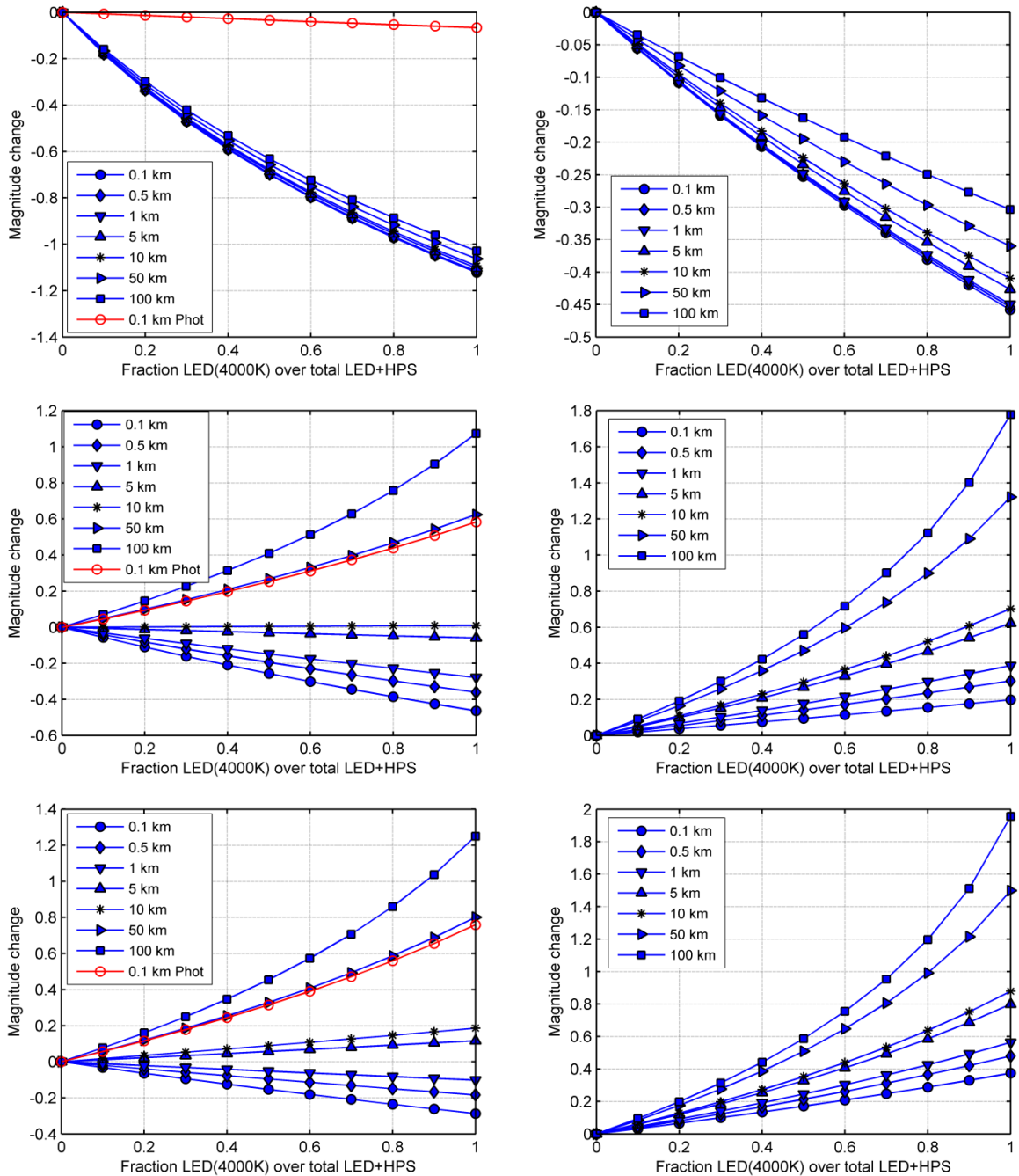

**Figure 6.** Evolution of the artificial zenithal night sky brightness in the visual (left column) and SQM (right column) photometric bands for the three transition scenarios. Upper row: Business as usual; Middle row: Zero direct upward flux; Lower row: Lighting reduction. See parameters in Table 3.

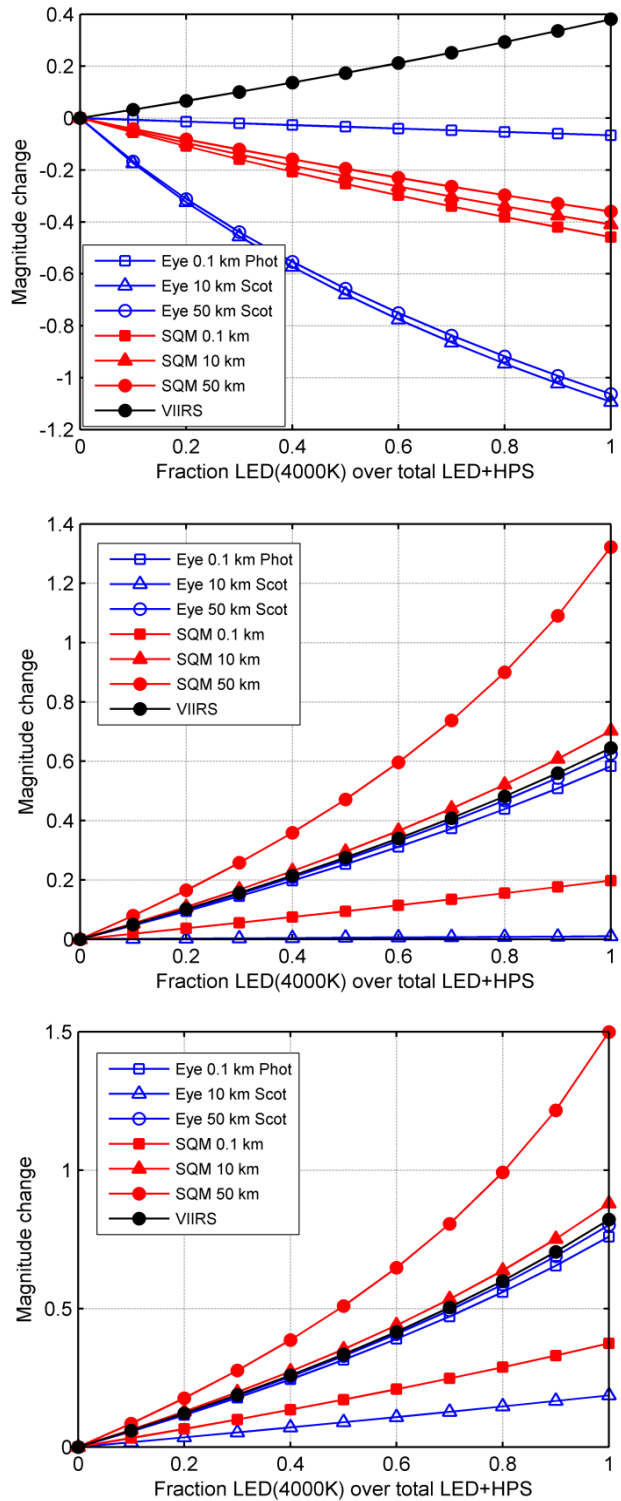

**Figure 7.** Combined plot of the evolution of the light pollution indicators (artificial zenithal night sky brightness at selected distances from the source, in the visual and SQM photometric bands, and VIIRS-DNB signal) for the three transition scenarios. Top: Business as usual; Middle: Zero direct upward flux; Bottom: Lighting reduction. See parameters in Table 3.

## 4. Discussion

The trends in Figures 6 and 7 are consistent with the prediction by Sánchez de Miguel et al (for the observation of the direct radiance of the lamps) [48] that the brightness evolution in the transition process from HPS to high CCT LED may show qualitatively different behaviors, depending on the observation band.

The detrimental role of the direct emissions to the upper hemisphere at angles close to the horizontal is readily apparent from Figures 6 and 7. If the transition from HPS to 4000 K LED is made by keeping constant these upward emissions at the $U_{LED} = U_{HPS} = 15\%$ level (Figure 6, upper row), the expected result is a noticeable increase of the zenithal sky brightness at all distances from the source, both in the visual and the SQM bands.

If the direct emissions of the LED luminaires to the upper hemisphere are cancelled ($U_{LED} = 0$, Figure 6, middle row), the zenithal sky is expected to get darker in the SQM band at all distances from the sources, as the transition process progresses. However, the visual appearance of the sky will behave in a different way, depending on the distances: for scotopically adapted observers at close distances from the sources (up to 5-10 km, depending on the scenario) it is expected that the HPS to 4000 K LED transition will lead to a brightening of the zenithal sky, due to the larger amount of scattered blue light from the LED lamps, in comparison with HPS ones. At large distances from the sources (> 10 km), in turn, the stronger attenuation of the blue components of the LED beams is expected to give rise to a net darkening of the zenithal sky for scotopic observers, in comparison with the full HPS starting point.

An interesting feature shown in Figure 7 is that along this transition process, and for all scenarios considered, the VIIRS-DNB radiometer is expected to detect a progressive reduction of the artificial light emissions, a trend not always accompanied by a similar reduction in the artificial zenithal sky brightness, either in the visual or in the SQM band. This is particularly conspicuous in case of a 'business as usual" transition (see Figure 7, top), where the sky brightness increases in all observation bands, and at all distances from the sources, as the HPS lit target area is progressively replaced by LED. The main reason for the behavior of the VIIRS-DNB signal is that when HPS lamps are replaced by 4000 K LED the radiance within the passband of the satellite radiometer (500-900 nm) is strongly reduced due to the loss of the near infra-red (NIR) HPS line around 814 nm, whereas the LED blue peak around 460 nm remains largely undetected.

These trends were deduced for a particular set of lamp spectra, radiance angular distribution, atmospheric conditions and detection photometric bands. It is expected, however, that the general traits described above can be representative of a certain range of situations of practical interest. According to additional calculations not shown here in detail, these trends are not substantially modified for other atmospheric compositions (e.g. $\Omega^{(A)}$=0.85, $g$=0.90, $\lambda_0$=500 nm, $\tau_0^{(A)}(\lambda_0)$=0.4, $\alpha$=1, see definitions of these parameters in Table 2), as well as for HPS lamps with zero NIR emissions, although in this last case the reduction of the VIIRS-DNB signal along the transition process is correspondingly less sharp.

Note that the photopic, scotopic, and SQM results displayed in Figures 6 and 7 correspond to the artificial component of the zenithal sky brightness. The natural brightness of the night sky, including moonlight when appropriate, shall be added to the artificial radiance $L_R(\lambda, t; D; \boldsymbol{\omega})$ in Equation (8) if the total radiance is to be determined. This total radiance can then be used in Equation (10) to determine the total signal finally recorded in the different channels.

Several limitations reduce the scope of this work. First, we have considered only two visual sensitivity functions, photopic (for very close distances to the sources < 0.5 km), and scotopic (for the whole distance range). Mesopic sensitivity would also be an appropriate metric for situations in which the adaptation luminance falls in between the two conditions. Second, only single scattering events have been included in the atmospheric propagation model. Whilst this approach can be reasonably accurate for small to medium propagation distances in low aerosol content atmospheres, it can lead to less correct results at large distances and/or under strong aerosol concentrations. The inclusion of double and multiple scattering effects [60] should be considered in future works. Third, a small source approximation (size << $D$) has been made to write Equation (8) from Equation (7). For sources of arbitrary size, the $K(\lambda; \boldsymbol{\omega}; D)$ function in Equation (7) can be used as a point spread function to compute the total spectral radiance, by spatially integrating this Equation over the whole source region.

## 5. Conclusions

In this work we have shown that the HPS to LED transition in outdoor lighting systems may give rise to different and even opposite behaviors in several commonly used light pollution indicators. The main reason are the combined effects of the changes in the sources' spectra, the wavelength-dependent atmospheric propagation processes, and the specific spectral sensitivity bands of the different detectors, including the human eye, the widely used SQM sensor and the VIIRS-DNB radiometer. Thus, in some transition scenarios, it is possible to detect a reduction of the emitted artificial radiance in the VIIRS-DNB band whereas the zenithal sky becomes visually brighter for scotopic observers located at short to medium distances from the sources, and darker for observers located at large distances. The zenithal sky brightness measured in the SQM band may show specific trends not necessarily coincident with the visual ones. The model described in this work allows to quantify these effects, under the simplifying assumption of single-scattering propagation in a layered atmosphere. The possible presence of this differential behavior should be taken into account when evaluating light pollution indicator datasets for assessing the outcomes of public policy decisions regarding the transformations of outdoor lighting systems.

**Acknowledgments**

This work was partially supported by Xunta de Galicia/FEDER, grant ED431B 2017/64. CITEUC is funded by National Funds through FCT - Foundation for Science and Technology (project: UID/MULTI/00611/2019) and FEDER - European Regional Development Fund through COMPETE 2020 – Operational Programme Competitiveness and Internationalization (project: POCI-01-0145-FEDER-006922).

**Appendix**

The steps leading to Equation 7 of the main text are briefly described in this appendix. To keep the model simple we restrict it to considering single-scattering processes in the molecules and aerosols that constitute the atmosphere. We also asume a layered atmosphere with properties only dependent on the height above ground. The source spectral radiance will be assumed to be dependent on the zenith angle, but azimuthally symmetric. Whereas azimuthal symmetry is not a feature that can be ascribed to most individual luminaires, it can be approximately fulfilled by ensembles of them at the city district level, excepting, perhaps, for urban nuclei with very marked street orientations. The main equations below closely follow those of Kocifaj [55], generalized to arbitrary, non Garstang, source radiances.

Following the general steps described in [55] one can show that the spectral radiance from the direction $\boldsymbol{\omega} = (z, \varphi)$, measured in the observer's reference frame, due to a source path of area $dA_0$ located at a distance $D$ from the observer, at an azimuth $\varphi_c$, is given by the integral:

$$dL_R(\lambda, t; \boldsymbol{\omega}) = dA_0 \int_{h=0}^{\infty} L(\lambda, t; z_{0,h}) \frac{\cos^3(z_{0,h})}{\cos(z)} \frac{T_\lambda(h, z, \varphi)}{h^2} \Gamma_\lambda(h, z, \varphi) dh \quad (A1)$$

where the integration is carried out over $h$, the height above ground level, with both source and observer located at the same baseline altitude. $L(\lambda, t; z_{0,h})$ is the spectral radiance of the azimuthally symmetric source, dependent on the zenith emission angle $z_{0,h}$, and generally variable in time. The subindex $h$ indicates that $z_{0,h}$ is the zenith angle, measured in the source reference frame, corresponding to the emitted rays that intercept the observer's line-of-sight at an altitude $h$ above the ground. This angle, that depends also on $D$ and $\varphi_c$, is given by [55]:

$$\cos(z_{0,h}) = \left\{1 + \tan^2 z + \frac{D}{h}\left[\frac{D}{h} - 2\tan z \cos(\varphi - \varphi_c)\right]\right\}^{-1/2} \quad (A2)$$

The term $T_\lambda(h, z, \varphi)$ within the integrand of (A1) accounts for the overall attenuation of the radiance, due to atmospheric extinction, along the combined paths from the source to the scattering volume located in $(z, \varphi, h)$, and from this volume to the observer. It is given by [55]:

$$T_\lambda(h, z, \varphi) = \exp\left\{\left[\frac{1}{\cos(z_{0,h})} + \frac{1}{\cos(z)}\right]\left[\tau_0^{(M)}(\lambda)\left[e^{-h/h_m} - 1\right]\right.\right.$$
$$\left.\left. + \tau_0^{(A)}(\lambda)\left[e^{-h/h_a} - 1\right]\right]\right\} \qquad (A3)$$

where $h_m$ and $h_a$ are the characteristic heights of the molecular and aerosol atmospheric concentration profiles, and $\tau_0^{(M)}(\lambda)$ and $\tau_0^{(A)}(\lambda)$ are the molecular and aerosol optical depths of the atmosphere, respectively. The molecular optical depth is given by $\tau_0^{(M)}(\lambda) = 0.00879\,\lambda^{-4.09}$, with $\lambda$ expressed in microns [61], and the aerosol optical depth is given by $\tau_0^{(A)}(\lambda) = \tau_0^{(A)}(\lambda_0)(\lambda/\lambda_0)^{-\alpha}$, where $\alpha$ is the Angstrom exponent, with both $\lambda$ and the arbitrary reference wavelength $\lambda_0$ expressed in homogeneous units (e.g. nm).

The term $\Gamma_\lambda(h, z, \varphi)$ contains information about the angular distribution of the single-scattered light [55], and it is itself the sum of two terms,

$$\Gamma_\lambda(h, z, \varphi) = \frac{\tau_0^{(M)}(\lambda)}{h_m}\exp\left\{-\frac{h}{h_m}\right\} \times \frac{3(1 + \cos^2(\vartheta))}{16\pi} +$$
$$\Omega^{(A)}\frac{\tau_0^{(A)}(\lambda)}{h_a}\exp\left\{-\frac{h}{h_a}\right\} \times \frac{1 - g^2}{4\pi(1 + g^2 - 2g\cos(\vartheta))^{3/2}} \qquad (A4)$$

the first one corresponding to Rayleigh molecular scattering and the second one to the aerosol contribution, assuming a Henyey-Greenstein angular phase function with asymmetry parameter $g$. $\Omega^{(A)}$ is the aerosol albedo, being the corresponding molecular one, $\Omega^{(M)}$, equal to 1. The scattering angle $\vartheta$, formed by the directions of propagation from the source to the elementary scattering volume and from this to the observer, is given by:

$$\cos(\vartheta) = \frac{1}{2}\left[\frac{D^2}{h^2}\cos(z)\cos(z_{0,h}) - \frac{\cos(z_{0,h})}{\cos(z)} - \frac{\cos(z)}{\cos(z_{0,h})}\right]. \qquad (A5)$$

The overall spectral flux $d\Phi_T(\lambda, t)$ emitted by the area $dA_0$ is

$$d\Phi_T(\lambda, t) = dA_0 \int_\Omega L(\lambda, t; z_{0,h})\cos(z_{0,h})d\omega, \qquad (A6)$$

where $d\omega = \sin(z_{0,h})\,dz_{0,h}d\phi$, and the integral is extended to the upper hemisphere $\Omega$ of size $2\pi$ steradians (sr).

Now, we can define the function $K(\lambda; \boldsymbol{\omega}; D)$ as the ratio of the spectral radiance in the observation direction to the spectral flux emitted by the source as

$$K(\lambda; \boldsymbol{\omega}; D) = \frac{dL_R(\lambda, t; \boldsymbol{\omega})}{d\Phi_T(\lambda, t)}, \qquad (A7)$$

from which Equation (7) immediately follows.

Two particular cases of the Equations (A1), (A6) and (A7) are considered in this work. First, the reflected radiance contribution $L_r(\lambda, t)$ in Equation (3) is assumed to be Lambertian, such that $L(\lambda, t; z_{0,h}) = L_r(\lambda, t)$, and this function, which does not depend on $h$, can be taken out of the integral in Equation (A1). For such a Lambertian source the emitted spectral flux, given by Equation (A6), is $d\Phi_T(\lambda, t) = dA_0 \pi L_r(\lambda, t)$. For the Lambertian component of the source emissions we have, then:

$$K_r(\lambda; \boldsymbol{\omega}; D) = \frac{1}{\pi} \int_{h=0}^{\infty} \frac{\cos^3(z_{0,h}) \, T_\lambda(h, z, \varphi)}{\cos(z)} \Gamma_\lambda(h, z, \varphi) dh \qquad (A8)$$

Second, the light directly emitted by the luminaires toward the upper hemisphere is assumed to be restricted to the zenithal directions comprised between $z_0 = \zeta$ and $z_0 = \pi/2$ (horizontal emission), expressed in the luminaires' reference frame. In this case, only those integration intervals $dh$ in Equation (A1) for which $z_{0,h}$ belongs to the interval $[\zeta, \pi/2]$ contribute effectively to the total result. The same happens in Equation (A6). Assuming a simplified model of constant radiance $L(\lambda, t; z_{0,h}) = L_u(\lambda, t)$ within this interval, and zero otherwise, we have $d\Phi_u(\lambda, t) = dA_0 L_u(\lambda, t) \pi[1 - \sin^2(\zeta)]$, and

$$K_u(\lambda; \boldsymbol{\omega}; D) = \frac{1}{\pi[1 - \sin^2(\zeta)]} \int_{h(\pi/2)}^{h(\zeta)} \frac{\cos^3(z_{0,h}) \, T_\lambda(h, z, \varphi)}{\cos(z)} \Gamma_\lambda(h, z, \varphi) dh \qquad (A9)$$

**References**

[1] Longcore T, Rich C. Ecological light pollution. Frontiers in Ecology and the Environment 2004;2:191-198.

[2] Rich C, Longcore T. Ecological consequences of artificial night lighting. 1st ed. Washington, D.C.: Island Press; 2004.

[3] Hölker F, Wolter C, Perkin EK, Tockner K. Light pollution as a biodiversity threat. Trends in Ecology and Evolution 2010;25:681-682.

[4] Davies TW, Bennie J, Inger R, Gaston KJ. Artificial light alters natural regimes of night-time sky brightness. Sci. Rep. 2013;3:1722. https://doi.org/10.1038/srep01722

[5] Gaston KJ, Bennie J, Davies TW, Hopkins J. The ecological impacts of nighttime light pollution: a mechanistic appraisal. Biological Reviews 2013;88:912–927.

[6] Gaston KJ, Duffy JP, Gaston S, Bennie J, Davies TW. Human alteration of natural light cycles: causes and ecological consequences. Oecologia 2014;176:917–931.

[7] Davies TW, Duffy JP, Bennie J, Gaston KJ. The nature, extent, and ecological implications of marine light pollution. Frontiers in Ecology and the Environment 2014;12(6): 347–355.

[8] Davies TW, Duffy JP, Bennie J, Gaston KJ. Stemming the Tide of Light Pollution Encroaching into Marine Protected Areas. Conservation Letters 2016;9(3):164–171. https://doi.org/10.1111/conl.12191.

[9] Longcore T, Rodríguez A, Witherington B, Penniman JF, Herf L, Herf M. Rapid assessment of lamp spectrum to quantify ecological effects of light at night. J Exp Zool. 2018;1–11. https://doi.org/10.1002/jez.2184

[10] Walker MF. The California Site Survey. Publ Astron Soc Pac 1970;82:672-698.

[11] Garstang R H. Night-sky brightness at observatories and sites. Publ Astron Soc Pac 1989;101:306-329.

[12] Cinzano P, Falchi F, Elvidge C. The first world atlas of the artificial night sky brightness. Mon Not R Astron Soc 2001;328:689–707.

[13] Falchi F, Cinzano P, Duriscoe D, Kyba CCM, Elvidge CD, Baugh K, Portnov BA, Rybnikova NA, Furgoni R. The new world atlas of artificial night sky brightness. Sci Adv 2016;2:e1600377 https://doi.org/10.1126/sciadv.1600377

[14] Marín C, Jafari J. StarLight: A Common Heritage. StarLight Initiative La Palma Biosphere Reserve, Instituto De Astrofísica De Canarias, Government of The Canary Islands, Spanish Ministry of The Environment, UNESCO-MaB: Canary Islands, Spain, 2008.

[15] Bogard P. The End of Night: Searching for Natural Darkness in an Age of Artificial Light. 1st ed. New York:Little, Brown and Company; 2013.

[16] Pauley SM. Lighting for the human circadian clock: recent research indicates that lighting has become a public health issue. Medical Hypotheses 2004;63:588–596.

[17] Duffy JF, Czeisler CA. Effect of light on human circadian physiology. Sleep Medicine Clinics 2009;4:165–177.

[18] Haim A, Portnov B. Light Pollution as a New Risk Factor for Human Breast and Prostate Cancers. Heidelberg: Springer; 2013. https://doi.org/10.1007/978-94-007-6220-6.

[19] Fonken LK, Nelson RJ. The Effects of Light at Night on Circadian Clocks and Metabolism. Endocrine Reviews 2014;35(4):648–670. https://doi.org/10.1210/er.2013-1051

[20] Russart KLG, Nelson RJ. Light at night as an environmental endocrine disruptor. Physiology & Behavior 2018;190:82–89. http://doi.org/10.1016/j.physbeh.2017.08.029

[21] Aubé M, Roby J. Sky brightness levels before and after the creation of the first International Dark Sky Reserve, Mont-Mégantic Observatory, Québec, Canada. Journal of Quantitative Spectroscopy & Radiative Transfer 2014;139:52–63.

[22] Dobler G, Ghandehari M, Koonin SE, Sharma MS. A hyperspectral survey of New York City lighting technology. Sensors 2016;16:2047.

[23] Alamús R, Bará S, Corbera J, Escofet J, Palà V, Pipia L, Tardà A. Ground-based hyperspectral analysis of the urban nightscape. ISPRS Journal of Photogrammetry and Remote Sensing 2017;124:16–26. http://doi.org/10.1016/j.isprsjprs.2016.12.004

[24] Kuechly HU, Kyba CCM, Ruhtz T, Lindemann C, Wolter C, Fischer J, Hölker F. Aerial survey and spatial analysis of sources of light pollution in Berlin, Germany. Remote Sensing of Environment 126;2012:39–50. http://dx.doi.org/10.1016/j.rse.2012.08.008

[25] Kyba CCM, Ruhtz T, Lindemann C, Fischer J, Hölker F. Two Camera System for Measurement of Urban Uplight. Angular Distribution Radiation Processes in the Atmosphere and Ocean (IRS2012) AIP Conf. Proc. 1531, 568-571 (2013). http://doi.org/10.1063/1.4804833

[26] Baugh K, Hsu FC, Elvidge CD, Zhizhin M. Nighttime Lights Compositing Using the VIIRS Day-Night Band: Preliminary Results, Proceedings of the Asia-Pacific Advanced Network 2013 v. 35, p. 70-86, (2013) http://doi.org/10.7125/APAN.35.8

[27] Kyba CCM, Garz S, Kuechly H, Sánchez de Miguel A, Zamorano J, Fischer J, Hölker F. High-Resolution Imagery of Earth at Night: New Sources, Opportunities and Challenges. Remote Sens 2015;7:1-23. http://doi.org/10.3390/rs70100001

[28] Estrada-García R, García-Gil M, Acosta L, Bará S, Sanchez de Miguel A, Zamorano J. Statistical modelling and satellite monitoring of upward light from public lighting. Lighting Research and Technology (2016) 48: 810-822. Published online, 21 April 2015. DOI: 10.1177/1477153515583181

[29] Elvidge CD, Baugh K, Zhizhin M, Hsu FC, Ghosh T. VIIRS night-time lights. International Journal of Remote Sensing 2017;38:5860-5879. http://doi.org/10.1080/01431161.2017.1342050

[30] Stefanov WL, Evans CA, Runco SK, Wilkinson MJ, Higgins MD, Willis K. Astronaut Photography: Handheld Camera Imagery from Low Earth Orbit, in J.N. Pelton et al. (eds.), Handbook of Satellite Applications, Springer International Publishing Switzerland, 2017. http://doi.org/10.1007/978-3-319-23386-4_39

[31] Zheng Q, Weng Q, Huang L, Wang K, Deng J, Jiang R, Ye Z, Gan M. A new source of multi-spectral high spatial resolution night-time light imagery—JL1-3B. Remote Sens Environ 2018;215:300–312.

[32] Pun CSJ, So CW. Night-sky brightness monitoring in Hong Kong. Environmental Monitoring and Assessment 2012;184(4):2537-2557. http://doi.org/10.1007/s10661-011-2136-1

[33] Puschnig J, Posch T, Uttenthaler S. Night sky photometry and spectroscopy performed at the Vienna University Observatory. Journal of Quantitative Spectroscopy & Radiative Transfer 2014;139:64–75.

[34] Puschnig J, Schwope A, Posch T, Schwarz R. The night sky brightness at Potsdam-Babelsberg including overcast and moonlit conditions. Journal of Quantitative Spectroscopy & Radiative Transfer 2014;139:76–81.

[35] Espey B, McCauley J. Initial Irish light pollution measurements and a new sky quality meter-based data logger. Lighting Research and Technology 2014;46(1):67-77.

[36] Kyba CCM, Tong KP, Bennie J, et al. Worldwide variations in artificial skyglow. Sci. Rep. , 2015;5:8409. http://doi.org/1010.1038/srep08409

[37] Bará S. Anthropogenic disruption of the night sky darkness in urban and rural areas. Royal Society Open Science 2016;3:160541. http://doi.org/10.1098/rsos.160541

[38] Hänel A, Posch T, Ribas SJ, Aubé M, Duriscoe D, Jechow A, Kollath Z, Lolkema DE, Moore C, Schmidt N, Spoelstra H, Wuchterl G, Kyba CCM. Measuring night sky brightness: methods and challenges. Journal of Quantitative Spectroscopy & Radiative Transfer 2018;205: 278–290. http://doi.org/10.1016/j.jqsrt.2017.09.008

[39] Posch T, Binder F, Puschnig J. Systematic measurements of the night sky brightness at 26 locations in Eastern Austria. Journal of Quantitative Spectroscopy & Radiative Transfer 2018;211:144–165. http://doi.org/10.1016/j.jqsrt.2018.03.010

[40] Rabaza O, Galadí-Enríquez D, Espín-Estrella A, Aznar-Dols F. All-Sky brightness monitoring of light pollution with astronomical methods. Journal of Environmental Management 2010;91:1278e1287.

[41] Duriscoe DM, Luginbuhl CB, Moore CA. Measuring night-sky brightness with wide-field CCD camera. Publ Astron Soc Pac 2007;119:192–213.

[42] Jechow A, Ribas SJ, Canal-Domingo R, Hölker F, Kolláth Z, Kyba CCM. Tracking the dynamics of skyglow with differential photometry using a digital camera with fisheye lens. Journal of Quantitative Spectroscopy & Radiative Transfer 2018;209:212-223.

[43] Jechow A, Hölker F, Kyba CCM. Using all-sky differential photometry to investigate how nocturnal clouds darken the night sky in rural areas. Scientific Reports 2019;9:1391. https://doi.org/10.1038/s41598-018-37817-8

[44] Kyba CCM, Wagner JM, Kuechly HU, Walker CE, Elvidge CD, Falchi F, Ruhtz T, Fischer J, Hölker F. Citizen Science Provides Valuable Data for Monitoring Global Night Sky Luminance. Sci Rep 2013;31835 https://doi.org/10.1038/srep01835

[45] Kyba CCM, Kuester T, Sánchez de Miguel A, Baugh K, Jechow A, Hölker F, Bennie J, Elvidge CD, Gaston KJ, Guanter L. Artificially lit surface of Earth at night increasing in radiance and extent. Sci. Adv. 2017;3:e1701528 https://doi.org/10.1126/sciadv.1701528

[46] Bertolo A, Binotto R, Ortolani S, Sapienza S. Measurements of Night Sky Brightness in the Veneto Region of Italy: Sky Quality Meter Network Results and Differential Photometry by Digital Single Lens Reflex. J. Imaging 2019;5:56. https://doi.org/10.3390/jimaging5050056

[47] Bará S, Lima RC, Zamorano J. Monitoring long-term trends in the anthropogenic brightness of the night sky. Sustainability 2019;11:3070 https://doi.org/10.3390/su11113070

[48] Sánchez de Miguel A, Aubé M, Zamorano J, Kocifaj M, Roby J, Tapia C. Sky Quality Meter measurements in a colour-changing world. Mon Not R Astron Soc 2017;467(3):2966-2979. https://doi.org/10.1093/mnras/stx145

[49] CIE, Commision Internationale de l'Éclairage. CIE 1988 2° SpectralLuminous Efficiency Function for Photopic Vision. Vienna: Bureau Central de la CIE; 1990.

[50] CIE, Commission Internationale de l'Éclairage. Recommended system for mesopic photometry based on visual performance. Pub. CIE 191:2010. Vienna: Bureau Central de la CIE; 2010.

[51] Pravettoni M, Strepparava D, Cereghetti N, Klett S, Andretta M, Steiger M. Indoor calibration of Sky Quality Meters: Linearity, spectral responsivity and uncertainty analysis. Journal of Quantitative Spectroscopy & Radiative Transfer 2016;181:74–86.

[52] Bará S, Tapia CE, Zamorano J. Absolute Radiometric Calibration of TESS-W and SQM Night Sky Brightness Sensors. Sensors 2019;19(6):1336. https://doi.org/10.3390/s19061336

[53] Miller S, Straka W, Mills S, Elvidge C, Lee T, Solbrig J, Walther A, Heidinger A, Weiss S. Illuminating the capabilities of the suomi national polar-orbiting partnership (npp) visible infrared imaging radiometer suite (viirs) day/night band. Remote Sensing 2013;5(12):67176766. http://dx.doi.org/10.3390/rs5126717

[54] Cao C, Bai Y. Quantitative Analysis of VIIRS DNB Nightlight Point Source for Light Power Estimation and Stability Monitoring, Remote Sens. 2014;6:11915-11935, doi:10.3390/rs61211915

[55] Kocifaj M. Light-pollution model for cloudy and cloudless night skies with ground-based light sources. Appl Opt 2007;46:3013-3022.

[56] Kocifaj M. A review of the theoretical and numerical approaches to modeling skyglow: Iterative approach to RTE, MSOS, and two-stream approximation, Journal of Quantitative Spectroscopy & Radiative Transfer 2016;181:2–10.

[57] Pogson N. Magnitude of 36 of the minor planets. Monthly Notices of the Royal Astronomy Society 1856;17:12.

[58] Kokaly RF, Clark RN, Swayze GA, Livo KE, Hoefen TM, Pearson NC, et al. USGS Spectral Library Version 7: U.S. Geological Survey Data Series 1035; 2017. https://doi.org/10.3133/ds1035.

[59] Fernández AJ, Molero F, Salvador P, Revuelta A, Becerril-Valle M, Gómez-Moreno FJ, et al. Aerosol optical, microphysical and radiative forcing properties during variable intensity African dust events in the Iberian Peninsula. Atmospheric Research 2017;196:129–141. http://dx.doi.org/10.1016/j.atmosres.2017.06.019

[60] Kocifaj M. Multiple scattering contribution to the diffuse light of a night sky: A model which embraces all orders of scattering, Journal of Quantitative Spectroscopy & Radiative Transfer 2018;206:260-272. https://doi.org/10.1016/j.jqsrt.2017.11.020

[61] Teillet PM. Rayleigh optical depth comparisons from various sources. Appl Opt 1990;29:1897-1900. https://doi.org/10.1364/AO.29.001897